\documentclass{desyproc}

\usepackage{epsfig,latexsym,cancel,amssymb,amsmath}
\usepackage{graphicx}

\DeclareMathOperator{\real}{Re}
\DeclareMathOperator{\imag}{Im}

\newcommand{\eV}{\ensuremath{\mathrm{eV}}}

\newcommand{\MeV}{\ensuremath{\mathrm{MeV}}}

\newcommand{\Uoneprime}{\ensuremath {\mathrm{U}(1)_V}}

\begin{document}
\title{New Light on Dark Photons}

\author{{\slshape Haipeng An$^1$, Maxim Pospelov$^{1,2}$, Josef Pradler$^{3}$%
    \footnote{Speaker. 9th Patras Workshop on Axions, WIMPs and
      WISPs.}
}\\[1ex]
$^1$Perimeter Institute for Theoretical Physics, Waterloo, ON N2L 2Y5, Canada\\
$^2$Department of Physics and Astronomy, University of Victoria,
Victoria, BC V8P 5C2, Canada $^3$Department of Physics and Astronomy,
Johns Hopkins University, Baltimore, MD 21218, USA}

\contribID{Pradler\_Josef}

\desyproc{DESY-PROC-2013-04}
\acronym{Patras 2013} 
\doi  

\maketitle

\begin{abstract}

  ``Dark Photons'', light new vector particles $V_{\mu}$ kinetically
  mixed with the photon, are a frequently considered extension of the
  Standard Model. For masses below 10~keV they are emitted from the
  solar interior. In the limit of small mass $m_V$ the dark photon
  flux is strongly peaked at low energies and we demonstrate that the
  constraint on the atomic ionization rate imposed by the results of
  the XENON10 Dark Matter experiment sets the to-date most stringent
  limit on the kinetic mixing parameter of this model: $\kappa \times
  m_V< 3\times10^{-12}\,\eV$. The result significantly improves
  previous experimental bounds and surpasses even the most stringent
  astrophysical and cosmological limits in a seven-decade-wide
  interval of $m_V$.
\end{abstract}

\section{Introduction}

In the recent years, the model of light vector particles with kinetic
mixing to the Standard Model photon has received tremendous attention,
theoretically as well as experimentally.
Whereas $m_{V} \gtrsim 1\,\MeV$ is mainly being probed in
medium-to-high energy collider experiments, masses in the sub-MeV
regime are subject to severe astrophysical and cosmological
constraints. Below $m_V < 10 $ eV, those limits are complemented by
direct laboratory searches for dark photons in non-accelerator type
experiments.  Among the most prominent are the
``light-shining-through-wall'' experiments (LSW)~\cite{Ahlers:2007qf}
and the conversion experiments from the solar dark photon flux,
``helioscopes''~\cite{Redondo:2008aa}; a collection of low-energy
constraints on dark photons can \textit{e.g.}~be found in the recent
review~\cite{Jaeckel:2010ni}.
Helioscopes derive their sensitivity from the fact that such light
vectors are easily produced in astrophysical environments, such as in
the solar interior, covering a wide range of masses up to $m_V \sim $
few keV. In general, stellar astrophysics provides stringent
constraints on any type of light, weakly-interacting particles once
the state becomes kinematically accessible~\cite{Raffelt:1996wa}. Only
in a handful of examples does the sensitivity of terrestrial
experiments match the stellar energy loss constraints.

Here we review our works~\cite{An:2013yfc,An:2013yua} in which we have
identified a new stellar energy loss mechanism originating from the
resonant production of longitudinally polarized dark photons and
derived ensuing constraints from underground rare event
searches. Limits on dark photons were improved to the extent that
previously derived constraints from all LSW and helioscope experiments
are now superseded by the revised astrophysical and new experimental
limits.

\section{Dark Photons from the sun: flux and detection}
\label{sec:main}

\begin{figure}[t]
\centering
\includegraphics[width=0.5\textwidth]{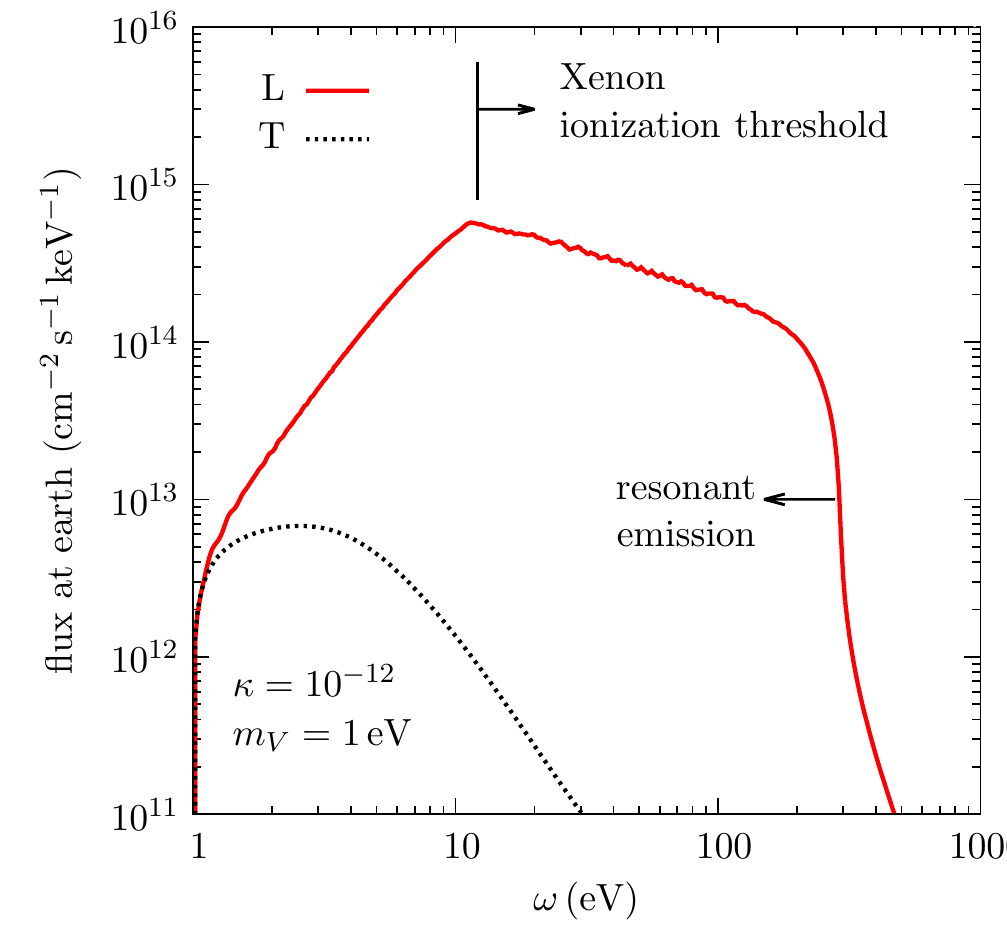}%
\includegraphics[width=0.5\textwidth]{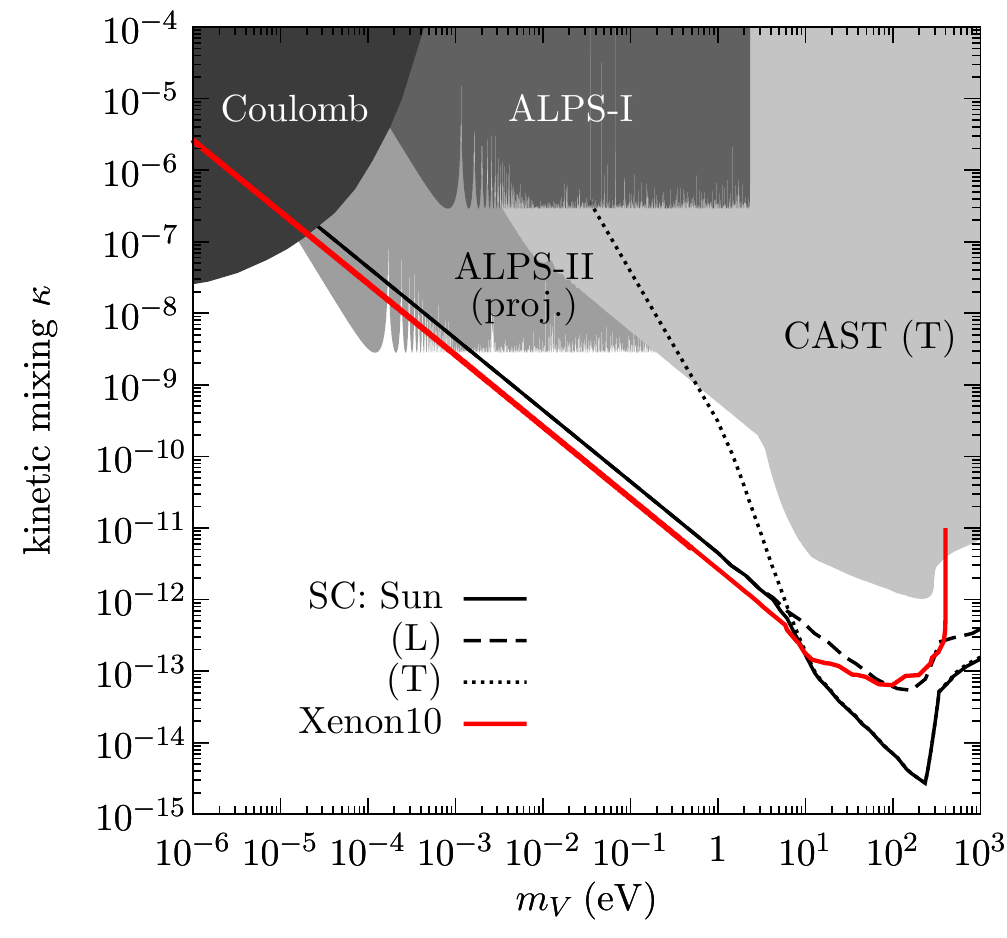}%
\caption{\small \textit{Left:} Energy differential solar dark photon
  flux at Earth for $\kappa = 10^{-12}$ and $m_V=1\,\eV$. The
  solid/dotted line shows the longitudinal(L)/transverse(T)
  contribution. \textit{Right:} Constraints on $\kappa$ as a function
  of $m_V$. The black solid/dashed/dotted curves show the
  total/longitudinal/transverse energy loss limit of the Sun by
  requiring that the dark photon luminosity does not exceed 10\% of
  the standard solar luminosity~\cite{Gondolo:2008dd}. The red line
  shows the constraint derived from the XENON10 data. Previous and
  future (=''proj.'') experimental bounds/sensitivities are shown by
  the shaded regions.  From light to dark shading these are from the
  CAST experiment~\cite{Andriamonje:2007ew} considering the
  contributions from only the transverse modes~\cite{Redondo:2008aa},
  from the ALPS collaboration~\cite{Ehret:2010mh}, and from tests of
  the inverse square law of the Coulomb  
  interaction~\cite{Bartlett:1988yy}.}
\label{fig:dp}
\end{figure}

The minimal extension of the SM gauge group by an additional
$\Uoneprime$ gauge factor yields the following effective Lagrangian
well below the electroweak scale,
\begin{align}
  \label{eq:L}
  \mathcal{L} = -\frac{1}{4} F_{\mu\nu}^2-\frac{1}{4} V_{\mu\nu}^2 -
  \frac{\kappa}{2} F_{\mu\nu}V^{\mu\nu} + \frac{m_V^2}{2} V_{\mu}V^{\mu}
  + e J_{\mathrm{em}}^{\mu} A_{\mu} ,
\end{align}
where $V_{\mu}$ is the vector field associated with the Abelian factor
$\Uoneprime$. The field strengths of the photon $F_{\mu\nu} $ and of
the dark photon $ V_{\mu\nu}$ are connected via the kinetic mixing
parameter $\kappa$ where a dependence on the weak mixing angle was
absorbed;  $ J_{\mathrm{em}}^{\mu}$ is the usual electromagnetic
current with electric charge~$e$.

Because of the U(1) nature of (\ref{eq:L}), we must distinguish two
cases for the origin of $m_V$: the Stueckelberg case (SC) with
non-dynamical mass, and the Higgs case (HC), where $m_V$ originates
through the spontaneous breaking of \Uoneprime\ by a new Higgs field
$h'$.  The crucial difference between the two cases comes in the small
$m_V$ limit: while all processes of production or absorption of $V$ in
SC are suppressed, $\Gamma_{\rm SC} \sim O(m_V^2)$, in HC there is no
decoupling, and $\Gamma_{\rm HC} \sim O(m_V^0)$.  Indeed, in the limit
$m_{V,h'}\to 0$ the interaction resembles one of a mini-charged scalar
with the effective EM charge of $e_{\rm eff} = \kappa
e'$~\cite{Holdom:1985ag,Okun:1982xi,Davidson:1991si,Davidson:2000hf}. In
the following we discuss the SC and refer the reader to our
work~\cite{An:2013yua} as well as to~\cite{Ahlers:2008qc} and
references therein for HC.

\paragraph{Solar flux}
\label{sec:flux}

The solar flux of dark photons in the SC is thoroughly calculated in
Ref.~\cite{An:2013yfc}; for further discussion see
also~\cite{Redondo:2013lna}. In the small mass region,
$m_V\ll\omega_p$ where $\omega_p$ is the plasma frequency, the
emission of longitudinal modes of $V$ dominates the total flux, and
the emission power of dark photons per volume can be approximated as
\begin{equation}
\label{res}
\frac{d P_L}{d V } \approx \frac{1}{4\pi} \frac{\kappa^2 m_V^2 \omega_p^3}{e^{\omega_p/T} - 1}  .
\end{equation} 
This formula is most readily obtained by noting that a resonant
conversion of longitudinal plasmons into dark photons is possible
whenever $\omega^2 = \omega^2_{p}$. The energy-differential flux of dark
photons at the location of the Earth is shown in the left panel of
Fig.~\ref{fig:dp}. Resonant emission stops for $\omega\gtrsim
300\,\eV$ since $\omega_p$ is limited by the temperature in the sun's
core.

\paragraph{Absorption of dark photons}
\label{sec:DDS}

In the SC, the ionization of an atom $A$ in the detector can then be
schematically described as $ V + A \to A^+ +e^- $.  The total dark
photon absorption rate is given by,
\begin{align}
\label{eq:Gamma}
\Gamma_{T,L} & = - \frac{\kappa_{T,L}^2 \imag{\Pi_{T,L}}}{\omega},
\qquad \kappa_{T,L}^{2} = \frac{\kappa^{2} m_{V}^{4}}{(m_{V}^{2} -
  \real \Pi_{T,L})^{2} + (\imag \Pi_{T,L})^{2}} .
\end{align}
$\kappa_{T,L}$ are the effective mixings for the transverse (T) and
longitudinal (L) modes respectively. The polarization functions
$\Pi_{T,L}$ are found from the in-medium polarization tensor
$\Pi^{\mu\nu}$,
\begin{align}
\label{eq:poltensor}
  \Pi^{\mu\nu}(q) = i e^2 \int d^4x\, e^{i q\cdot x} \langle \Omega |
  T J^{\mu}_{em} (x) J^{\nu}_{em} (0) | \Omega \rangle = - \Pi_T
  \sum_{i=1,2} \varepsilon_i^{T\mu} \varepsilon_i^{T\nu} -
  \Pi_L\varepsilon^{L\mu} \varepsilon^{L\nu} ,
\end{align}
where $q=(\omega, \vec q)$ is the dark photon four momentum and
$\epsilon_{\mu}^{T,L}$ are the polarization vectors for the transverse
and longitudinal modes of the dark photon, $\epsilon_\mu^2=-1$. The
first relation (\ref{eq:Gamma}) is a manifestation of the optical
theorem.

The polarization functions $\Pi_{T,L}$ are related to the complex
index of refraction, $n_{\rm refr}$ or, equivalently, to the
permittivity of the medium $\varepsilon = n_{\mathrm{refr}}^2 $.  For
an isotropic, non-magnetic medium $ \quad \Pi_{L} = (\omega^2 - \vec
q^2) (1- n_{\mathrm{refr}}^2),$ and $ \Pi_T = \omega^2 (1-
n_{\mathrm{refr}}^2)$, so that for an incoming on-shell dark photon
with $q^2=m_V^2$, $\Gamma_L \propto \kappa^2 m_V^2$ indeed holds.
We obtain $ n_{\mathrm{refr}} $ from its relation to the forward
scattering amplitude $f(0) = f_1 + i f_2$ where the atomic scattering
factors $f_{1,2}$ are
\textit{e.g.}~tabulated in~\cite{Henke:1993eda}. Close to the ionization
threshold we make use of the Kramers-Kronig dispersion relations to
relate $f_1$ and $f_2$ for estimating $n_{\rm refr}$. Alternatively,
one can solve an integral equation relating $\imag \varepsilon$ and
$\real \varepsilon$ in a self-consistent manner, an approach taken
in~\cite{An:2013yua}.

\paragraph{Limits from direct detection}
\label{sec:DDS}

With flux $d\Phi_{T,L}/d\omega$ and absorption rate $\Gamma_{T,L}$ at hand, the
expected number of signal events in a given experiment reads
\begin{equation}\label{master1}
N_{\rm exp} = V T \int^{\omega_{\rm max}}_{\omega_{\rm min}} \frac{\omega d \omega}{|\vec{q}|} \left(\frac{d\Phi_{T}}{d\omega} \Gamma_T + \frac{d\Phi_{L}}{d\omega} \Gamma_L\right) {\rm Br}  ,
\end{equation}
where $V$ and $T$ are the fiducial volume and live time of the
experiment, respectively, and ${\rm Br}$ is the branching ratio of
photoionization rate to total absorption rate.

Given the significant infrared enhancement of the solar dark photon
spectrum, left panel of Fig.~\ref{fig:dp}, the low-energy ionization
signals measured in the XENON10~\cite{Angle:2011th} dark matter
experiment have the best sensitivity to constrain a dark photon flux
that is also supported by the Sun.  With $\sim$12~eV ionization energy in
xenon, the absorption of a dark photon with 300~eV energy can produce
about 25 electrons.
From~\cite{Angle:2011th} we estimate a 90\% C.L upper limit on the
detecting rate to be $r < 19.3$ events kg$^{-1}$day$^{-1}$ (similar to
limits deduced in Ref. \cite{Essig:2012yx}).  In the region $12{~\rm
  eV}<\omega<300$~eV the ionization process dominates the absorption,
and therefore ${\rm Br}$ in this region can be set to unity. The 90\%
C.L. upper limit on $\kappa$ as a function of $m_V$ is shown by the
thick red curve in Fig.~\ref{fig:dp}. As can be seen it surpasses
other current experimental limits as well as the solar energy loss
bound in a mass interval from $10^{-5}\,\eV < m_V \lesssim 10\,\eV$.

Given the enormous amount of experimental progress in the field of
direct Dark Matter detection, one can be optimistic that future
sensitivity to dark photons, and other light particles is bound to be
further improved.

\section{Acknowledgments}

The speaker would like to thank the conference organizers for financial
support.


\begin{footnotesize}

\end{footnotesize}


\end{document}